\documentclass[aps,nofootinbib,twocolumn,]{revtex4}

\usepackage{amsmath,amsfonts}

\newcommand{\bea}{\begin{eqnarray}}
\newcommand{\eea}{\end{eqnarray}}
\def\beann{\begin{eqnarray*}}
\def\eeann{\end{eqnarray*}}

\newcommand{\beq}{\begin{equation}}
\newcommand{\eeq}{\end{equation}}

\newcommand{\ba}{\begin{array}}
\newcommand{\ea}{\end{array}}

\def\ben{\begin{enumerate}}
\def\een{\end{enumerate}}

\def\4{\tilde }
\def\5{\bar }
\def\6{\partial }
\def\7{\hat }

\def\cF{{\cal F}}
\def\cB{{\cal B}}

\def\cG{{\cal G}}
\def\cC{{\cal C}}

\def\cL{{\cal L}}

\def\cG{{\cal G}}
\font\mybb=msbm10 at 10pt
\def\bb#1{\hbox{\mybb#1}}
\def\bR {\bb{R}}

\begin{document}

\onecolumngrid
\begin{flushright}
   WIS/06/03-June-DPP\\
   hep-th/0306057
\end{flushright}

\title{G\"odel's Universe in a Supertube Shroud}
\author{
Nadav Drukker, Bartomeu Fiol and Joan Sim\'{o}n}
\email{drukker, fiol, jsimon@weizmann.ac.il}
\affiliation{Department of Particle Physics,
The Weizmann Institute of Science,
2 Herzl Street, Rehovot 76100, Israel}


\begin{abstract}
We demonstrate that certain supersymmetric G\"odel-like universe solutions 
of supergravity are not 
solutions of string theory. This is achieved by realizing that supertubes
are BPS states in these spaces, and under certain conditions, 
when wrapping closed timelike curves, some world-volume modes develop negative
kinetic terms. Since these universes are
homogeneous, this instability takes place everywhere in space-time. 
We also construct a family of supergravity solutions which locally look like 
the G\"odel universe inside a domain wall made out of supertubes, but 
have very different asymptotic structure. One can adjust the volume inside 
the domain wall so there will be no closed timelike curves, and then those 
spaces seem like perfectly good string backgrounds.
\end{abstract}

\maketitle


\section{Introduction.}

Einstein's equations in General Relativity have many solutions that seem 
unphysical. Some have curvature singularities, 
which generically lie beyond the regime of validity of the classical theory.
Other solutions may have no singularities, but violate causality, by 
having closed timelike curves.
Hawking \cite{hawking} has presented arguments to support the Chronology 
Protection Conjecture: namely that when quantum mechanics is taken into 
account, backgrounds not having closed timelike curves do not develop them.
Even if this is 
true, it seems to leave open the question of whether backgrounds having 
closed timelike curves to begin with have to be discarded or not.

The answer to such questions may not be found entirely within classical 
general relativity. One would like to know that the energy momentum tensor, 
serving as the source in Einstein's equations corresponds to reasonable 
physical matter. Another direction to address the problem is to ask 
whether there is a way to make sense of physics on spaces with closed 
timelike curves in classical mechanics, quantum mechanics or quantum 
field theory.

Probably best of all is to ask those questions in string theory, which 
encompasses both general relativity and quantum physics.
Just as string theory has managed to resolve certain curvature singularities,
mainly timelike, by the addition of extra degrees of freedom, it is natural
to wonder about the status of closed timelike curves in such a quantum 
theory of gravity.
Are such backgrounds allowed in string theory? Does string theory realize 
the Chronology Protection conjecture? Some recent
work in this direction is 
\cite{horava,dyson,Jarv:2002wu,Gimon:2003ms,Herdeiro:2002ft,Biswas:2003ku}.

In string theory, the existence of dynamical extended objects provides us with
probes suited particularly well to study non-local issues like closed timelike 
curves. Instead of trying to quantize the string worldsheet
theory in one of these backgrounds, finding its spectrum and adding 
interactions, our strategy will be 
to look for BPS states corresponding to extended branes in this vacuum and to
analyze the dynamics of small fluctuations around them. For a specific subset
of backgrounds having closed timelike curves we will find BPS states with 
a sickness in their effective action, signaling the invalidity of the 
solution.

Let us be more precise. There are many supergravity backgrounds having 
closed timelike curves.
We will be interested in G\"odel-like universes, which were first embedded
in five dimensional supergravity \cite{gauntlett}. Some other examples of
similar metrics were found in \cite{horava,harmark}.  All these
solutions are generalizations of a four dimensional metric studied by
Som and Raychaudhuri \cite{somray}, and are very similar to the
original G\"odel universe \cite{godel}.

All these solutions describe rotating spaces, 
where one can choose the planes of rotation as well 
as some fluxes that make the solution supersymmetric. We will concentrate on a 
specific solution of type IIA supergravity with rotation in a single plane, 
since this is the simplest example of such a metric with closed timelike 
curves.

This solution \cite{harmark} has a nontrivial metric in three
dimensions, which we parameterize by the time coordinate $t$, and
polar coordinates $r$ and $\phi$ in the plane. There is a NS-NS
flux as well as RR 2-form and 4-form fluxes, and we label
the extra direction in which there is flux by $y$. The metric and fluxes
are
\begin{equation}
  \begin{gathered}
    ds^2 = -\left(dt+cr^2d\phi\right)^2+dr^2+r^2d\phi^2+dy^2 
      +\sum_{i=4}^9 (dx^i)^2 \\
    H_3=-2cr\,dr\wedge d\phi\wedge dy\,,\qquad F_2=-2cr\,dr\wedge d\phi\,,
    \\ F_4=2cr\,dt\wedge dr\wedge d\phi\wedge dy\,.
  \end{gathered}
 \label{srmetric}
\end{equation}
This one parameter solution preserves 8 supercharges and we shall assume
positive $c$ without loss of generality. If we consider motion in the periodic
$\phi$ direction at constant $r$, the corresponding spacetime curve becomes
timelike for $r>1/c$, hence the closed timelike curves.

Though this seems like a perfectly valid solution to the equations of
supergravity, we will demonstrate that, as it stands, it can't be a valid
solution of string theory. One way of realizing this fact is to notice that
supertubes \cite{mateos,emparan} can exist in the above background.
Supertubes are bound states of D0-branes and fundamental strings
having non-vanishing D2-dipole moment. They can
be realized on the worldvolume of a cylindrical D2-brane with electric and
magnetic fields turned on, such that they induce a non-vanishing angular
momentum that balances the tension which would naturally tend to collapse
the tube. In our discussion, supertubes extend in the $y$ direction and
wrap the angular direction $\phi$ at fixed radius $r$. It is not too
surprising that this is the most interesting object to probe the above
geometry with, 
since it has the correct rotational symmetry, and couples to all the
background fields that were turned on. As will also be made 
clear below, this object preserves the same supersymmetries as the
G\"odel-like universe itself. Even though the tension of this supertube is
just given by the sum of the D0-brane and fundamental string charges,
and it can have an arbitrary size (by adjusting these charges), we will find 
that for a certain range of charges, some world-volume modes develop a 
negative kinetic term whenever the radius satisfies
$r\geq 1/c$, so that the supertube wraps a closed timelike curve.
This sickness is much like that found by considering a probe brane in the
repulson background, which led to the enhan\c{c}on mechanism
\cite{enhancon}. We interpret it as an instability of the vacuum itself,
which indicates that these supertubes will want to condense.

The plan of the rest of the paper is the following.  In the next
section we shall briefly discuss the probe calculation of a supertube in 
our G\"odel universe, and show that the world-volume theory may become sick
at the radius where closed timelike curves appear. 
In Section 3, we describe a family of supergravity solutions, that locally 
look like the G\"odel universe, but have a domain wall made out of smeared 
supertubes separating them from a space that does not have closed timelike 
curves asymptotically. The domain wall in these solutions can be thought of 
as a regulator, the initial solution (\ref{srmetric}) arising in the limit 
where the domain wall is sent to infinity. One might think it possible to 
enlarge the region inside the domain-wall to get closed timelike curves, but 
the same problem we encountered in the full G\"odel universe prevents this 
from happening.

We conclude with some discussion of the analogy to the enhan\c{c}on 
mechanism and how similar problems show up for other G\"odel universes of
supergravity, invalidating them too.

\section{Supertubes in a G\"odel-like universe.}

As mentioned in the introduction, the type IIA Som-Raychaudhuri
solution (\ref{srmetric}) has closed timelike curves. Thus, one may
suspect that an extended probe would develop some sickness in this
background. Our first goal is then to discuss the possible probes in
this background. This task is simplified by the realization that the
solution (\ref{srmetric}) is a particular case of a family of IIA solutions
discussed in \cite{emparan,paul}. The authors of \cite{emparan} prove that in
general these solutions admit D0 branes, fundamental strings and supertubes as 
probes that don't break any further supersymmetry. Even better, they perform a 
probe analysis of these solutions, and show that when the solution presents 
closed timelike curves, the worldvolume theory on the supertube is ill-defined.
Our computation is just a special case of theirs.

The supertube is a cylindrical D2-brane which is extended in the $y$ 
direction as well as the angular direction $\phi$ at fixed radius $r$ about 
the origin.\footnote{The metric (\ref{srmetric}) is actually homogeneous. The
addition of the supertube breaks translational invariance, but preserves
the rotation around the origin.} The world-volume theory on the supertube 
is just that of a D2-brane in curved background, which includes the
Dirac-Born-Infeld and Wess-Zumino terms
\bea
S=-\int e^{-\Phi}\sqrt{-\det(\cG+\cF)}
-\int(\cC_3+\cC_1\wedge\cF)\,,
\eea
where $\cG$ is the pullback of the metric and $\cC_1$ and $\cC_3$ the
pullbacks of the RR potentials. The field strength $\cF$ includes
the gauge fields that are turned on in the brane and those induced by
the NS field
\bea
\cF=F-B_2=E\,dt\wedge dy+\cB\,dy\wedge d\phi\,.
\eea
$E$ is the electric field and $\cB=B-cr^2$ the magnetic field, including
the term induced by the background.

The equations of motion are solved by $E=1$ (which is not critical in the 
presence of magnetic field) and constant\footnote{The positivity
of the magnetic field is enforced by supersymmetry.} $B>0$, whereas the 
radius of
the tube is fixed to be the angular momentum $J$, related to the D0 and
fundamental string charge densities $q_0=B$ and $q_s$ by
$r^2=J=q_0q_s$. It is a simple exercise to show that the supertube 
preserves the same supersymmetries as the background, which is also 
clear from the analysis of the next section, where the metric 
(\ref{srmetric}) is shown to belong to the family of metrics studied 
in \cite{emparan}.

After adjusting the electric and magnetic fields as above, the
Lagrangian density of the D-brane probe is simply the magnetic field
$\cL=-B$. The supertube tension is just given by the 
sum of charges it carries, that of D0-branes and fundamental strings, 
as dictated by supersymmetry. Therefore we can seemingly make the tube 
as small or as large as we wish.

To verify that, we can start with a very small supertube and slowly 
increase the radius. There are a number of ways one could envision this
happening. For instance, one could imagine a bath of D0-branes and fundamental
strings surrounding the supertube, and allowing the supertube to change its
radius by exchanging charges with the bath, satisfying the BPS condition at all
times. Perhaps more naturally, one can also consider fluctuations of the 
radius, in time and/or in the $y$ direction, which don't preserve the BPS 
condition. In this latter case, a non-zero potential would also appear.

This calculation was done in \cite{emparan} for the background
generated by a collection of supertubes, and is trivial to adapt it
to our metric. The result is that the expansion of the Lagrangian at small
velocities is
\bea
\cL = \cL_0+\frac{1}{2}M\,\dot r^2+\dots
\eea
where the dots can contain a potential term if the fluctuation doesn't preserve
the BPS condition. $\cL _0$ is the static Lagrangian density and
\bea
M=\frac{(B-cr^2)^2 + r^2-c^2r^4}{B}\,,
\eea
is the mass of the excited mode. This is positive definite 
for $r<1/c$, but for larger radii, it becomes negative for a certain range of 
charges.

How do we interpret this singularity? In similar computations, e.g.
when we 
encounter that the metric in moduli space becomes negative we interpret it 
as meaning that the effective Lagrangian is not valid, because it is missing 
some light degrees of freedom. One then corrects the effective description by 
adding those extra degrees of freedom. An example of this is the enhan\c {c}on
mechanism \cite{enhancon}. It is not exactly clear how to apply this 
procedure to the case at hand. Recall that the G\"odel universe 
(\ref{srmetric}) is homogeneous, so there are closed timelike curves through 
every point in space, and the sickness we found is not confined to a bounded
region of spacetime. In addition we don't have well defined 
asymptotic conditions, or well defined 
global charges that a new solution should reproduce (since it is not clear 
how to define charge in G\"odel universes). What we can safely conclude is 
that the solution (\ref{srmetric}) is not a good string theory background.

\section{A domain wall solution.}

In the previous section we argued that the G\"odel-type solution 
(\ref{srmetric})
can't be a correct effective description of a string theory background.
Nevertheless it is fairly easy to display supergravity solutions that
locally are similar to G\"odel, but are free of closed timelike curves.
To do so, we start by recalling the family of IIA solutions considered
in \cite{paul,emparan}.
\bea
  \begin{aligned}
 \label{supertube}
    ds^2=&-U^{-1}V^{-1/2}(dt-A)^2+U^{-1}V^{1/2}dy^2 \\
      &+V^{1/2}\sum_{i=2}^9 (dx^i)^2 \\
    B_2=&-U^{-1}(dt-A)\wedge dy+dt\wedge dy\,, \\
    C_1=&-V^{-1}(dt-A)+dt\,, \\
    C_3=&-U^{-1}dt\wedge dy\wedge A\,, \\
    e^\Phi=& U^{-1/2}V^{3/4}\,.
  \end{aligned}
\eea
Here $U$ and $V$ are harmonic functions in the eight dimensions
spanned by $x^i$, and $A$ is a Maxwell field. Clearly the IIA
G\"odel like solution (\ref{srmetric}) is recovered if we take $U=V=1$ and
$A=-cr^2d\phi$ (where $r$ and $\phi$ are polar coordinates in the
$x^2$, $x^3$ plane).

A very natural question is whether there are more general configurations 
which are G\"odel near the origin, but are asymptotically different, with
no closed timelike curves.\footnote{The idea of patching another
metric outside a finite radius, to construct a solution free of closed 
timelike curves, was 
considered for the original G\"odel metric in \cite{bonnor}.}. If we want 
to retain translation symmetry in $\bR^6$ and rotational symmetry 
in $\phi$, any source must be smeared in those directions, so $U$ and $V$ 
are harmonic functions on the plane with rotational symmetry.

We construct the solution by taking $U=V=1$ and $A=-cr^2d\phi$ near the
origin. At a radius $R$ we put the smeared supertube, and for $r> R$
we choose
\bea
&&    U=1+\frac{Q_s}{2\pi}\ln\frac{r}{R}\,,\qquad
      V=1+\frac{Q_0}{2\pi}\ln\frac{r}{R}\,,
      \nonumber\\
&&    A=-cR^2d\phi\,.
 \label{metout}
\eea
$Q_s$ and $Q_0$ are respectively the fundamental string and 
D0-brane charge densities, and are still arbitrary. The choice of $A$
is motivated by requiring continuity of the metric through 
the domain wall.

This space is identical to G\"odel near the origin, but has very
different asymptotics. For example, it has finite angular momentum, whereas 
for the original space it diverges. Also, inside the domain wall 
there are 
constant magnetic fields for the D0-brane and fundamental string, while 
outside there are mainly electric fluxes (the rotation of space mixes 
electric and magnetic fields).
One bad feature is that the dilaton grows as $r\rightarrow \infty$.

It is easily verified that the solution satisfies the Israel matching
conditions \cite{israel}, with the supertube stress-energy tensor at the
junction. Furthermore, the matching fixes the D2-brane density in terms of
$c$. A similar calculation was done for the enhan\c{c}on in \cite{myers}.

There is a very good analog to these solutions in
electromagnetism---a charged solenoid. Outside there is
electric flux, with the potential behaving like a log. 
The current around the
solenoid will induce some magnetic flux inside, but no electric fields.

From this construction it is also clear that the supertube preserves
the same supersymmetries as the G\"odel metric (\ref{srmetric}).
After all, the latter is just the metric induced by a large collection
of the same supertubes, but this can also be verified directly by the
same calculations as in \cite{paul,emparan,quim}.

So far we did not specify the radius where the domain wall is located. 
Clearly when $R\leq1/c$ there are no closed timelike curves within 
the domain wall, but it can also be shown that there are no such 
curves outside the shell. At most, if $R=1/c$ there are closed null 
curves on the shell itself, which might warrant further study.

If we place the shell at a larger radius there will be causality 
violating curves. It is therefore natural 
to think of the full G\"odel solution as the limit when the domain wall 
is taken to infinity and the region outside discarded. This picture may 
provide a better laboratory for studying the problems of the G\"odel 
universe. In particular, if the radius is only slightly larger than 
$1/c$, there is only a thin shell of closed timelike curves. So the 
sickness is not spread over the entire space, and it should be possible 
to describe a process by which the shell contracts to eliminate the 
region with closed timelike curves.

In any event, it seems like those spaces cannot be created. A
calculation similar to the one carried out in section II will show that, 
if one tries to increase the size of the region inside the shell, so it 
exceeds $R=1/c$, the same problems will appear, indicating that one cannot 
increase the radius of the shell further.

\section{Discussion}

One of the most striking uses of D-branes in string theory is as probes of
spacetime, often exposing the limitations of supergravity, and providing a
cure to some of its pathological solutions. The enhan\c{c}on mechanism 
\cite{enhancon}  provides a beautiful example of this. Studying supertubes 
on the G\"odel-like universe (\ref{srmetric}) teaches us a similar lesson, 
that the naive supergravity solution is not valid.
As opposed to the enhan\c{c}on case, we find that the space is sick 
everywhere, and the question of how to ``cure'' it does not seem well posed.

Nevertheless, the domain wall solutions we discussed in section III locally
look like G\"odel universes, and present certain analogies with the
enhan\c{c}on mechanism: in both cases, we keep part of the original metric,
and replace the problematic part (the repulson singularity in one case,
the region with closed timelike curves in the other) by another solution,
separated by a shell formed by the probes themselves. Of course, an obvious
difference is that in the enhan\c{c}on case one keeps
the asymptotic form of the metric and changes the interior, while in
our case we do the opposite. 

In this paper we concentrated on a specific G\"odel universe solution 
of supergravity. Other solutions were studied in 
\cite{gauntlett,horava,harmark} involving rotations in more planes. 
It is a simple check to see that all the solutions in 
type IIA and IIB suffer from the same problems we described. One has 
to take the same supertube, or in some cases the T-dual objects, and place them
so that their circles follow the closed timelike curves, and the 
rest of the calculation is identical. Therefore all those spaces are not 
good string theory backgrounds.
We expect this to be a general mechanism that eliminates many solutions with 
closed timelike curves in string theory.

\section*{Acknowledgements}

We are very grateful to Ofer Aharony, Tom Banks, Micha Berkooz, 
Jacques Distler, Roberto Emparan, Willy Fischler, Surya Ganguli, 
Shinji Hirano, Eliezer Rabinovici and Kip Thorne for
interesting discussions on related topics. We are particularly grateful 
to Roberto Emparan for pointing out a mistake in an earlier version 
of this paper.
We would also like to thank the ITP at
Stanford university and the organizers of the Stanford-Weizmann
workshop, where this project started. 
BF would further like to thank the Santa Cruz 
Institute for Particle Physics at UCSC and 
JS would like to thank the Perimeter Institute and ITP at the University of 
Amsterdam for their hospitality. 
BF is supported through a European 
Community Marie Curie Fellowship, and also by the Israel-U.S. Binational 
Science Foundation, the IRF Centers of Excellence program, the European RTN 
network HPRN-CT-2000-00122 and by Minerva.
JS is supported by the Phil Zacharia Postdoctoral Fellowship.



\begin{thebibliography}{99}




\bibitem{hawking}
S.~W.~Hawking,
Phys.\ Rev.\ D {\bf 46}, 603 (1992).


\bibitem{horava}
E.~K.~Boyda, S.~Ganguli, P.~Ho\v{r}ava and U.~Varadarajan,
Phys.\ Rev.\ D {\bf 67}, 106003 (2003)
[arXiv:hep-th/0212087].


\bibitem{dyson}
L.~Dyson,
arXiv:hep-th/0302052.


\bibitem{Jarv:2002wu}
L.~J\"arv and C.~V.~Johnson,
Phys.\ Rev.\ D {\bf 67}, 066003 (2003)
[arXiv:hep-th/0211097].


\bibitem{Gimon:2003ms}
E.~G.~Gimon and A.~Hashimoto,
arXiv:hep-th/0304181.


\bibitem{Herdeiro:2002ft}
C.~A.~Herdeiro,
arXiv:hep-th/0212002.


\bibitem{Biswas:2003ku}
R.~Biswas, E.~Keski-Vakkuri, R.~G.~Leigh, S.~Nowling and E.~Sharpe,
arXiv:hep-th/0304241.


\bibitem{gauntlett}
J.~P.~Gauntlett, J.~B.~Gutowski, C.~M.~Hull, S.~Pakis and H.~S.~Reall,
arXiv:hep-th/0209114.


\bibitem{harmark}
T.~Harmark and T.~Takayanagi,
arXiv:hep-th/0301206.


\bibitem{somray}
M.~M.~Som and A.~K.~Raychaudhuri
Proc. R. Soc. London A{\bf 304}, 81 (1968)


\bibitem{godel}
K.~G\"odel,
Rev.\ Mod.\ Phys.\  {\bf 21}, 447 (1949).


\bibitem{mateos}
D.~Mateos and P.~K.~Townsend,
Phys.\ Rev.\ Lett.\  {\bf 87}, 011602 (2001)
[arXiv:hep-th/0103030].


\bibitem{emparan}
R.~Emparan, D.~Mateos and P.~K.~Townsend,
JHEP {\bf 0107}, 011 (2001)
[arXiv:hep-th/0106012].

\bibitem{enhancon}
C.~V.~Johnson, A.~W.~Peet and J.~Polchinski,
Phys.\ Rev.\ D {\bf 61}, 086001 (2000)
[arXiv:hep-th/9911161].

\bibitem{paul}
J.~P.~Gauntlett, R.~C.~Myers and P.~K.~Townsend,
Phys.\ Rev.\ D {\bf 59}, 025001 (1999)
[arXiv:hep-th/9809065].


\bibitem{bonnor}
W.~B.~Bonnor, N.~O.~Santos and M.~A.~MacCallum,
Class.\ Quant.\ Grav. {\bf 15}, 357 (1998)
[arXiv:gr-qc/9711011].


\bibitem{israel}
W.~Israel,
Nuovo Cim.\ B {\bf 44S10} (1966) 1
[Erratum-ibid.\ B {\bf 48} (1967\ NUCIA,B44,1.1966) 463].


\bibitem{myers}
C.~V.~Johnson, R.~C.~Myers, A.~W.~Peet and S.~F.~Ross,
Phys.\ Rev.\ D {\bf 64}, 106001 (2001)
[arXiv:hep-th/0105077].

\bibitem{quim}
J.~Gomis, T.~Mateos, P.J.~Silva and A.~Van Proeyen,
arXiv:hep-th/0304210.



\end{thebibliography}
\end{document}